\documentclass{PoS}

\title{LO/NLO, LO* and jet algorithms}

\ShortTitle{LO/NLO, LO* and jet algorithms}

\author{\speaker{J. Huston}%
         \thanks{I would like to thank Zvi Bern, John Campbell, Lance Dixon, Daniel Maitre, Jan Winter and Giulia Zanderighi for useful conversations. I would like to dedicate this contribution to the memory of my friend Thomas Binoth, who I last saw at RadCor. }\\
        Michigan State University, Department of Physics and Astronomy, East Lansing, Michigan\\
        E-mail: \email{huston@msu.edu}}
\abstract{The impact of NLO corrections, and in particular the role of jet algorithms,  is examined for a  variety of processes at the Tevatron and LHC.}

\FullConference{RADCOR 2009 - 9th International Symposium on Radiative Corrections (Applications of Quantum Field Theory to Phenomenology) \\
		 October 25-30 2009\\
		 Ascona, Switzerland}

\begin{document}
\section{Introduction}
Next-to-leading order (NLO) is the first order at which the normalization, and in some cases the shape, of perturbative cross sections can be considered reliable~\cite{Campbell:2006wx}. A great deal of effort has recently been devoted towards the calculation of complex cross-sections at NLO. A prioritized list of NLO cross sections was assembled at Les Houches in 2005~\cite{Buttar:2006zd} and added to in 2007~\cite{Bern:2008ef}. This list includes cross sections which are experimentally important, and which are theoretically feasible (if difficult) to calculate. Basically all $2\rightarrow3$ cross sections of interest have been calculated, with the frontier now extending to $2\rightarrow4$ calculations. Often these calculations exist only as private codes. To reach full utility, the codes should be made public and/or the authors should generate ROOT ntuples providing the parton level event information from which experimentalists can assemble any cross sections of interest. Of course the ultimate goal will be the ability to link any NLO calculation to a parton shower Monte Carlo~\cite{NLOLHA}. 

\section{K-factors}
Experimentalists typically deal with leading order (LO) calculations, especially in the context of parton shower Monte Carlos. Some of the information from a NLO calculation can be encapsulated in the K-factor, the ratio of the NLO to LO cross section for a given process, with the caveat that the value of the K-factor depends upon a number of variables, including the values of the  renormalization and factorization scales, as well as the parton distribution functions (PDFs) used at LO and NLO. In addition, the NLO corrections often result in a shape change, so that one K-factor is not sufficient to describe the impact of the NLO corrections on the LO cross section.  
Even with these caveats, it is still useful to calculate the K-factors for interesting processes at the Tevatron and LHC. A K-factor table, originally shown in the CHS review article~\cite{Campbell:2006wx} and then later expanded in the Les Houches 2007 proceedings~\cite{Bern:2008ef}, is shown below. The K-factors are shown for several different choices of scale and with the use of either LO or NLO PDFs for the LO calculation. Also shown are the K-factors when the CTEQ modified LO PDFs are used~\cite{Lai:2009ne}. 
\begin{table}[h]
\begin{center}
\begin{tabular}{|l|l|l|c|c|c|c|c|c|c|}
\hline
  & \multicolumn{2}{|l|}{Fact. scales} & 
 \multicolumn{3}{|c|}{Tevatron K-factor} &
 \multicolumn{4}{|c|}{LHC K-factor} \\ 
  & \multicolumn{2}{|l|}{\quad} & 
 \multicolumn{3}{|c|}{} & \multicolumn{4}{|c|}{} \\
Process & $\mu_0$ & $\mu_1$ &
 ${\cal K}(\mu_0)$ & ${\cal K}(\mu_1)$ & ${\cal K}^\prime(\mu_0)$ &
 ${\cal K}(\mu_0)$ & ${\cal K}(\mu_1)$ & ${\cal K}^\prime(\mu_0)$ &
 ${\cal K}^{\prime\prime}(\mu_0)$  \\
\hline
&&&&&&&&&\\
$W$        & $m_W$ & $2m_W$	&
 1.33 & 1.31 & 1.21 & 1.15 & 1.05 & 1.15 & 0.95 \\
$W$+1 jet          & $m_W$ & $ p_T^{\rm jet}$ &
 1.42 & 1.20 & 1.43 & 1.21 & 1.32 & 1.42 & 0.99\\
$W$+2 jets & $m_W$ & $ p_T^{\rm jet}$	    &
 1.16 & 0.91 & 1.29 & 0.89 & 0.88 & 1.10 & 0.90\\
$WW$+1 jet                & $m_W$ & $2m_W$    &
 1.19 & 1.37 & 1.26 & 1.33 & 1.40 & 1.42 & 1.10\\
$t{\bar t}$        & $m_t$ & $2m_t$         &
 1.08 & 1.31 & 1.24 & 1.40 & 1.59 & 1.19 & 1.09\\
$t{\bar t}$+1 jet        & $m_t$ & $2m_t$    &
 1.13 & 1.43 & 1.37 & 0.97 & 1.29 & 1.10 & 0.85\\
$b{\bar b}$        & $m_b$ & $2m_b$         &
 1.20 & 1.21 & 2.10 & 0.98 & 0.84 & 2.51 & -- \\
Higgs      & $m_H$ & $ p_T^{\rm jet}$       & 
 2.33 & -- & 2.33 & 1.72 & -- & 2.32 & 1.43 \\
Higgs via VBF      & $m_H$ & $ p_T^{\rm jet}$ &
 1.07 & 0.97 & 1.07 & 1.23 & 1.34 & 0.85 & 0.83  \\
Higgs+1 jet     & $m_H$ & $ p_T^{\rm jet}$ &
 2.02 & -- & 2.13 & 1.47 & -- & 1.90 & 1.33\\
Higgs+2 jets     & $m_H$ & $ p_T^{\rm jet}$
 & -- & -- & -- & 1.15 & -- & -- & 1.13 \\
&&&&&&&&&\\
\hline
\end{tabular}
\caption{\label{tab:K-fact}
K-factors for various processes at the LHC (at 14 TeV) calculated using
a selection of input parameters.
In all cases, for NLO calculations, the CTEQ6M PDF set is used.
For LO calculations, ${\cal K}$ uses the CTEQ6L1 set,
whilst ${\cal K}^\prime$ uses the same PDF set, CTEQ6M,
as at NLO, and ${\cal K}^{\prime\prime}$ uses
the LO-MC (2-loop) PDF set CT09MC2.
For Higgs+1 or 2 jets, a jet cut of $40\ \mathrm{GeV}/c$
and $|\eta|<4.5$ has been applied. 
A cut of $p_{T}^{\mathrm{jet}}>20\ \mathrm{GeV}/c$
has been applied to the $t\bar{t}$+jet process,
and a cut of $p_{T}^{\mathrm{jet}}>50\ \mathrm{GeV}/c$ to the $WW$+jet
process. In the $W$(Higgs)+2 jets process, the jets are separated by $\Delta R>0.4$ 
(with $R_{sep}=1.3$), whilst the vector boson fusion (VBF) 
calculations are performed for a Higgs boson of mass $120$~GeV.
In each case the value of the K-factor is compared
at two often-used scale choices, $\mu_0$ and $\mu_1$.}
\end{center}
\end{table}

Several patterns can be observed in the K-factor table. NLO corrections appear to be larger for processes in which there is a great deal of color annihilation, such as $gg->Higgs$ in which two color octet gluons produce a color singlet Higgs boson. NLO corrections also tend to decrease as more final-state legs are added~\footnote{A rule-of-thumb derived by Lance Dixon is that the K-factor is often proportional to the factor $C_{i1} + C_{i2}  Ð -  C_{f,max}$, where $C_{i1}$ and $C_{i2}$ are the Casimir color factors for the initial state and $C_{f,max}$ is the Casimir factor for the biggest color representation that the final state can be in. Of course, this is not intended to be a rigorous rule, just an illustrative one.}. The K-factors at the LHC are similar to the K-factors for the same processes at the Tevatron, but have a tendency to be smaller. 
\section{W + 3 jets}
\begin{figure}
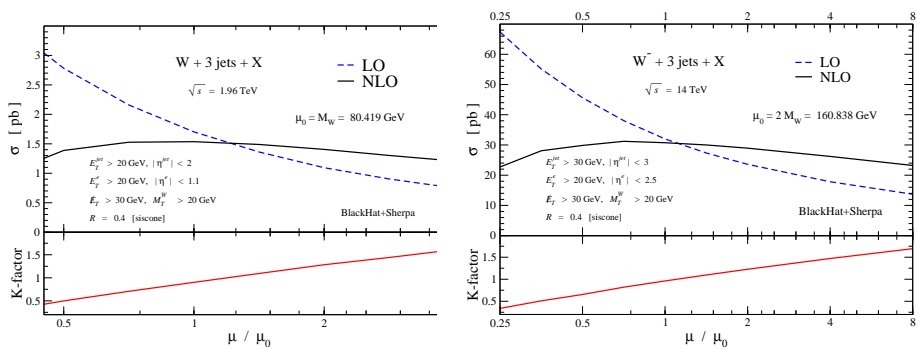

\begin{center}
\includegraphics[width=6cm,angle=0]{W3jTev_FullColor_scale_dependence.eps}
\includegraphics[width=6cm,angle=0]{Wm3jLHC_FullColor_scale_dependence.eps}
\end{center}
\vspace*{-0.5cm}
\caption{The scale dependence of the  cross sections for $W$ + 3 jet production at the Tevatron and LHC (14 TeV)~\cite{Berger:2009ep}.
\label{fig:scale_dep}}
\end{figure}

The cross section for the production a W boson and 3 jets has recently been calculated at NLO~\cite{Berger:2009ep}, \cite{KeithEllis:2009bu}. The scale dependence for this cross section is shown in Figure~\ref{fig:scale_dep} for the Tevatron and for the LHC(14 TeV)~\cite{Berger:2009ep}. It can be observed that, using a scale of $m_W$, the K-factor at the Tevatron is approximately unity, while at the LHC it less than 0.6. 

\begin{figure}[ht]
\includegraphics[width=0.8\columnwidth,height=1.6\textheight,keepaspectratio]
{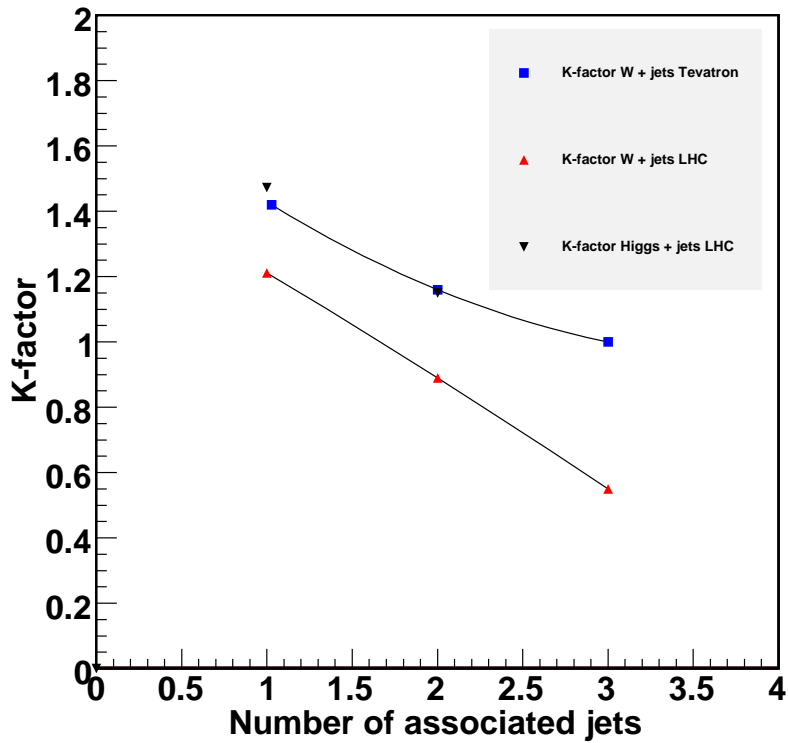}
\caption{The K-factors (NLO/LO) are plotted for $W$ production at the Tevatron and LHC and for Higgs production at the LHC as a function of the number of accompanying jets. The $k_T$ jet algorithm with a D parameter of 0.4 has been used. 
\label{fig:Kfact}}
\end{figure}

The K-factors for W + 1, 2 or 3 jets, at a renormalization/factorization scale  of $m_W$, are plotted in Figure \ref{fig:Kfact} (along with similar K-factors for Higgs + 1 or 2 jets)~\footnote{For these plots, the NLO CTEQ6.6 PDFs~\cite{Nadolsky:2008zw}
 have been used with both the LO and NLO matrix elements, in order to separate any PDF effects from matrix element effects. If a LO PDF such as CTEQ6L1 were used instead, the LO cross sections would shift upwards, but the trends would be the same.}. In this plot, a pattern becomes obvious. The K-factors appear to decrease linearly as the number of final state jets increases, with a similar slope at the Tevatron as at the LHC (but with an offset). A similar slope is observed for Higgs  boson+ jets at the LHC. To further  understand this pattern (in addition to the color flow argument discussed in the previous section), we first have to review jet algorithms at LO and NLO. 
At LO, one parton equals one jet. By choosing a jet algorithm with size parameter D, we are requiring any two partons to be a distance D or greater apart. The matrix elements have $1/\Delta R$ poles, so a larger value of D means smaller cross sections. At NLO, there can be two partons in a jet, and jets for the first time can have some structure. No $\Delta R$ cut is needed since the virtual corrections cancel the collinear singularity from the gluon emission (but there are residual logs that can become important if the value of D is too small). Increasing the size parameter D increases the phase space for including an extra gluon in the jet, and thus increases the cross section at NLO (in most cases). The larger the number of final state partons, the greater the differences will be between the LO and NLO dependence on jet size. 

\begin{figure}
\begin{center}
\includegraphics[width=7cm,angle=0]{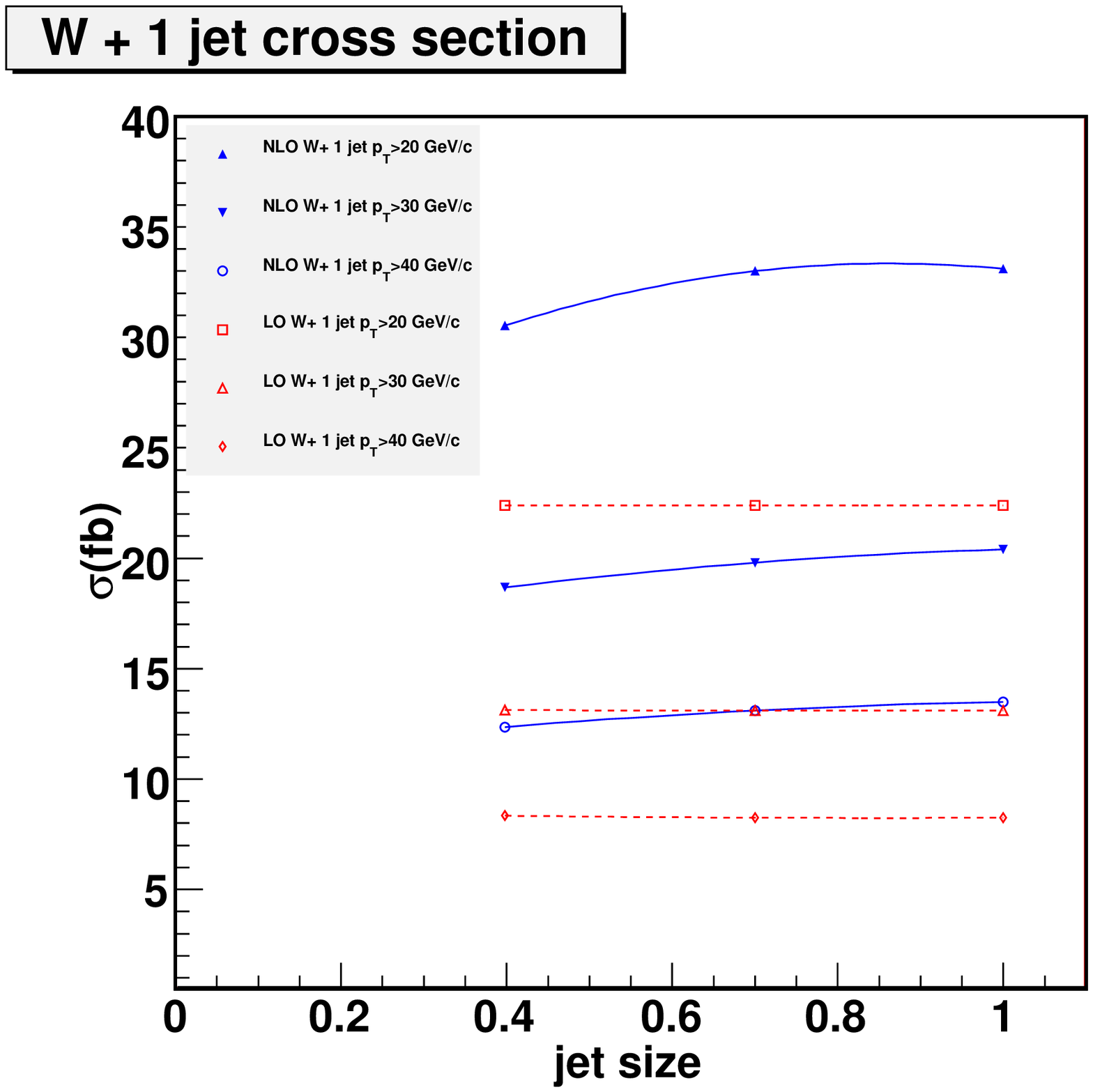}
\includegraphics[width=7cm,angle=0]{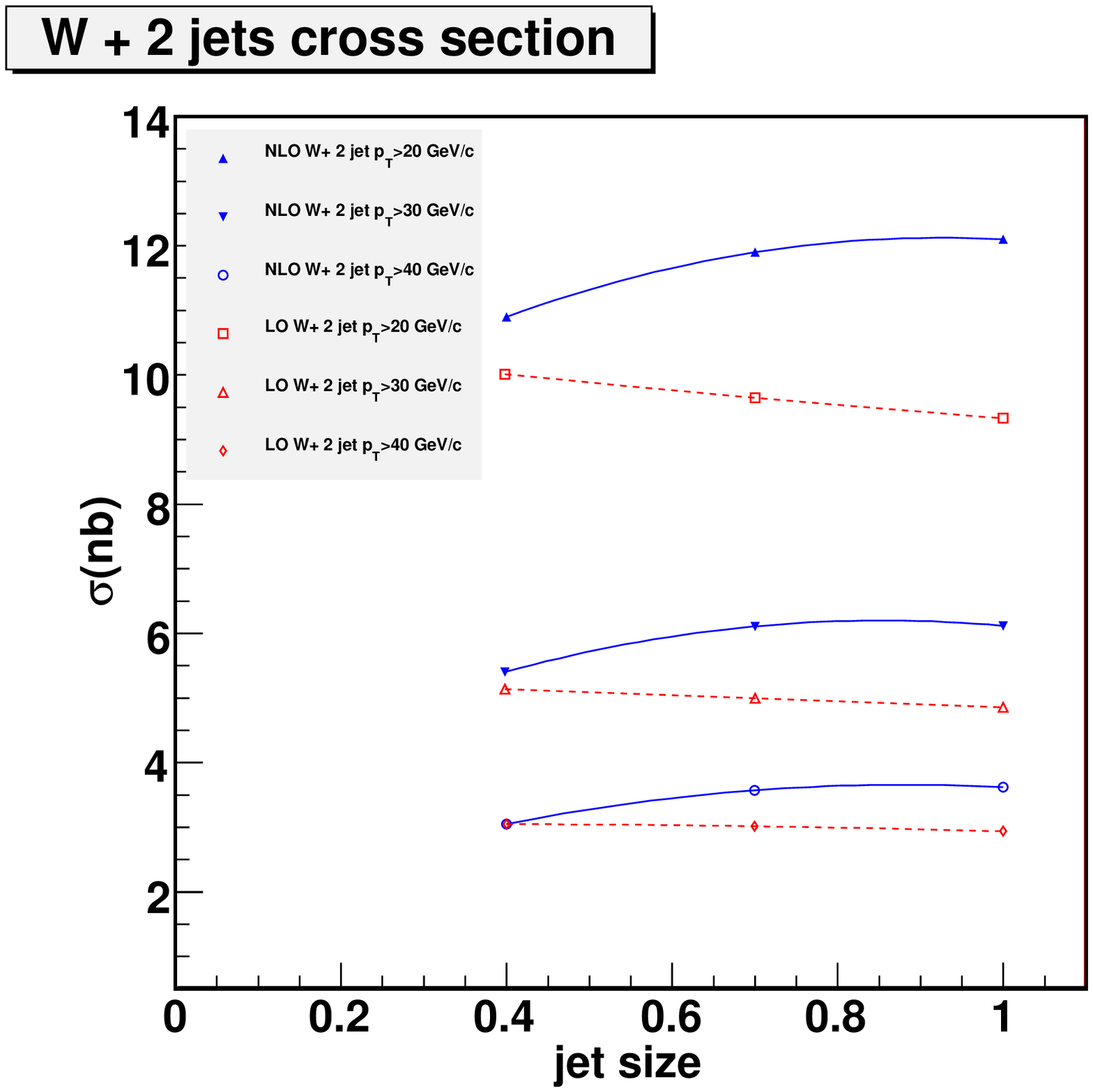}
\includegraphics[width=7cm,angle=0]{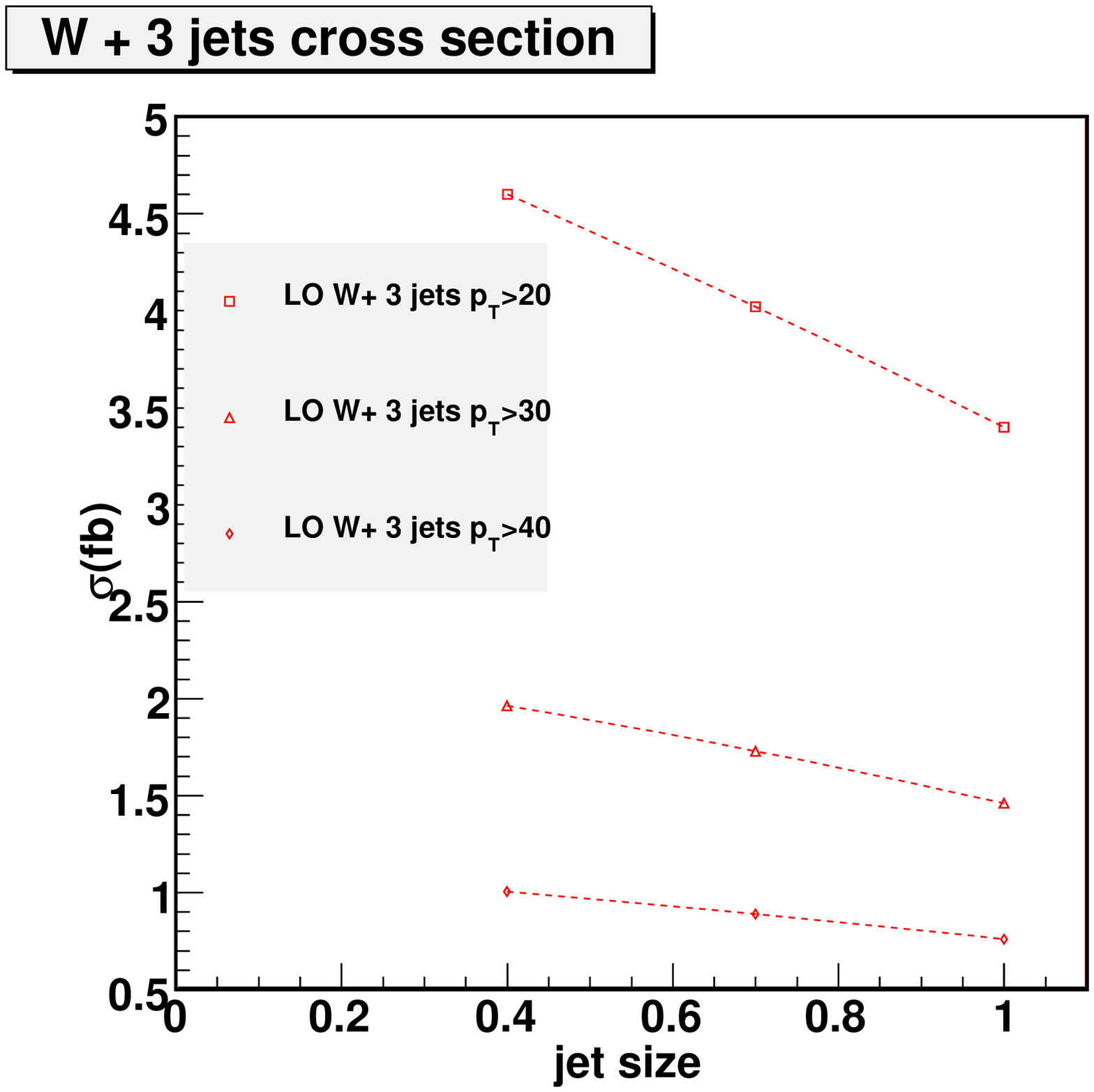}
\end{center}
\vspace*{-0.5cm}
\caption{The cross sections for W + 1, 2 and 3 jets at the LHC at LO and NLO (for the 1 and 2 jet case) as a function of the jet size (using the $k_T$ algorithm and of the jet transverse momentum. 
\label{fig:w_123}}
\end{figure}

In Figure~\ref{fig:w_123}, the cross sections for $W$ + 1, 2 and 3 jets are plotted as a function of the jet size and of the jet transverse momentum, at LO and NLO (for the 1 and 2 jet case). The NLO cross sections are observed to increase with increasing jet sizes, while the LO cross sections decrease (except for the trivial behavior for $W$ + 1 jet, where there is only 1 parton in the final state. The slope for the LO cross sections becomes steeper as the number of partons increases. NLO predictions for $W$ + 3 jets are not available, but would be very interesting to plot for comparison.  

In this context, the K-factor for W + 3 jets, at a scale of $m_W$, can be at least partially understood. The problem does not lie with the NLO cross section. That is well-behaved. The problem is that the LO cross section sits $\it too-high$, due at least partially to the collinear enhancement that comes from a small jet size (0.4). For soft gluons (on the order of 20 GeV/c), there is in addition a residual impact from a soft singularity. The K-factor for W + 3 jets at the LHC would be smaller if (a) a larger jet size were used (b) a larger jet transverse momentum were used, or (c)  a larger scale were used. In Ref.\cite{Berger:2009ep}, it has been shown that a scale such as $H_T$ results (at the LHC) not only in a K-factor closer to unity for $W$ + 3 jets, but in similar shapes for kinematic distributions at LO and NLO. Scales that are typically used at the Tevatron, such as $m_W$ or $m_W^2+p_{T,W}^2$ lead to low normalizations and kinematic shapes that can be significantly different at LO than at NLO. Another study (Ref. \cite{Melnikov:2009wh}) has found that the kinematic shapes at LO and NLO can also be made similar if a $\it local$ scale, such as that obtained with the CKKW~\cite{Catani:2001cc} procedure is used. The connection between these two observations is not obvious and deserves further investigation. 

\section{Jet Sizes}
From the experimental perspective, in complex final states such as $W$ + n jets, it is useful to have smaller jet sizes so as to be able to resolve the n jet structure. Smaller jet sizes can also reduce the impact of pileup and underlying event~\cite{Dasgupta:2007wa}. 
From the theoretical perspective:
\begin{itemize}
\item hadronization effects become larger as R decreases
\item for small R, the ln R perturbative terms referred to previously can become noticeable
\item the restriction in phase space for small R can affect the scale dependence, i.e the scale uncertainty for an n-jet final state can depend on the jet size
\end{itemize}
The jet sizes to be used at the LHC should depend primarily on the needs of the experimental analyses. However, it will still be important to understand the impact that any choice of a jet size may have on the LO and NLO predictions, and the relation between the two predictions. This is another motivation for the use of multiple jet algorithms (and parameters) at the LHC in order to fully understand/explore the wide range of QCD dynamics for both standard model and beyond the standard model physics~\cite{Ellis:2007ib}. 

\section{Conclusions}
The technology for NLO calculations has progressed to the point where  {$2\rightarrow4$} processes are being completed. In this contribution, I have examined the impact of the jet algorithms applied to the LO and NLO components of the multi-parton matrix elements and have demonstrated the tendency for the K-factor to decrease as the number of jets in the final state increases.

\end{document}